\documentclass{aa}
\usepackage{graphicx}
\usepackage{dcolumn}
\usepackage{bm}
\usepackage{natbib}
\usepackage{mathptmx}      
\usepackage{amsmath}

\begin{document}

\title{Orbit determination for next generation space clocks}

\author{Lo\"\i c Duchayne \inst{1}
\and Flavien Mercier \inst{2} \and Peter Wolf \inst{1}}

\institute{LNE-SYRTE, Observatoire de Paris, CNRS, UPMC \and Centre National d'Etudes Spatiales}

\date{\today}

\abstract {In this paper, we study the requirements on orbit determination compatible with operation of next generation space clocks at their
expected uncertainty. Using the ACES (Atomic Clock Ensemble in Space) mission as an example, we develop a relativistic model for time and frequency
transfer to investigate the effects of orbit determination errors. We show that, for the considered orbit error models, the required uncertainty goal
can be reached with relatively modest constraints on the orbit determination of the space clock, significantly less stringent than expected from
"naive" estimates. Our results are generic to all space clocks and represent a significant step towards the generalized use of next generation space
clocks in fundamental physics, geodesy, and time/frequency metrology.}


 \maketitle

\section{Introduction}
\label{intro}

Over the last decade of the 20$^{th}$ century and the first few years of the 21$^{st}$, the uncertainty of atomic clocks has decreased by over two
orders of magnitude, passing from the low $10^{-14}$ to below $10^{-16}$, in relative frequency \cite{Bize,Heavner,Oskay,Rosenband}. This rapid
evolution is essentially due to recent technological breakthroughs (laser cooling and trapping of atoms and ions), which allow very effective control
and reduction of the motion of the atoms and correspondingly long interrogation times. Atomic fountain microwave clocks use freely falling laser
cooled atoms and were the first to reach uncertainties below $10^{-15}$ some of them being at present in the low $10^{-16}$ range. They are now
closely followed, and even surpassed, by trapped ion and neutral atom optical clocks, the best of which show uncertainties below $10^{-16}$.

Space applications in fundamental physics, geodesy, time and
frequency metrology, navigation, etc... are among the most
promising for this new generation of clocks. Onboard terrestrial
or solar system satellites, their exceptional frequency stability
and accuracy make them a prime tool to test the fundamental laws
of nature, and to study the Earth's and solar system gravitational
potential and its evolution. In the longer term, they are likely
to provide the primary time reference for the Earth, as clocks on
the ground will be subject to the less accurate knowledge of the
geopotential on the surface \cite{Wolf1}.

For example, when comparing a clock on a low Earth orbiting
satellite ($\approx$~1000~km altitude) to one on the ground they
display a difference of $\approx 10^{-10}$ in relative frequency
due to the relativistic gravitational frequency shift. Measuring
that difference with $10^{-17}$ uncertainty would allow a test of
the gravitational frequency shift to a few parts in $10^7$ or
equivalently, a determination of the potential difference between
the clocks at the 10~cm level. The latter would contribute
significantly to the knowledge of the geopotential and related
applications in geophysics, representing the first realisation of
relativistic geodesy \cite{Bjerhammar,Soffel} where the
fundamental observable is directly the gravitational potential via
the relativistic redshift.

From the above it is obvious that next generation space clocks at the envisaged uncertainty level require a fully relativistic analysis and
modelling, not only of the clocks (in space and on the ground) but also of the time/frequency transfer method used to compare them
\cite{Allan,Klioner,Petit,Wolf1,Blanchet,Teyssandier}. Indeed, a highly accurate space clock is of limited use unless it can be compared to ground
clocks using a method that does not degrade the overall uncertainty, and unless the behaviour of the clocks as a function of their positions and
velocities can be modelled with sufficient accuracy. As an example, simple order of magnitude estimates of the relativistic gravitational frequency
shift show that an 1~m error on the position of the clocks leads to an error of $\approx 10^{-16}$ in the determination of their frequency
difference. Similarly when using an one-way system (GPS like) for the time transfer an 1~m position error leads to an error of $\approx
3\times10^{-9}$ s in the synchronisation ie. $\geq 10^{-14}$ in relative frequency over one day.

In this paper, we study in more detail the requirements on orbit determination compatible with operation of next generation space clocks at the
required uncertainty, and based on a completely relativistic model. We use the example of the ACES (Atomic Clock Ensemble in Space) mission, an
ESA-CNES project to be installed onboard the ISS (International Space Station) in 2013. It consists of two atomic clocks and a two-way time transfer
system (microwave link, MWL) with an overall uncertainty goal of 1 part in $10^{16}$ after ten day integration (see section \ref{relat} for more
details). We show that the required accuracy goal can be reached with relatively modest constraints on the orbit determination of the space clock of
$\approx 10$~m in position (for the considered orbit error models), which is about an order of magnitude less stringent than expected from "na\"\i
ve" estimates ($\approx 1$~m, see above). This is due to first order cancellation between the velocity and position part of the orbit determination
error in the determination of the relativistic frequency shift of the space clock, and to the use of a two-way time transfer system (MWL) which leads
to first order cancellation of the position errors in the clock comparison (see section \ref{results}). Our results are generic to all space clocks
(not limited to the ACES mission) and represent a significant step towards the generalised use of next generation space clocks in fundamental
physics, geodesy, and time/frequency metrology, as they show that the constraints on the orbit determination of the space clock are significantly
less stringent than previously thought.

In sections \ref{ACES} and \ref{relat} we briefly describe the ACES mission and the relativistic model used for the clocks and the time transfer. We
explicitly derive the effect of orbit errors on the clock comparison in section \ref{Effects}. Up to this point, our results are completely general
(within the specified approximations) with no assumptions on the expected orbit determination errors. We then apply those results using two examples
of expected orbit errors (section \ref{orbito}). Our main results are the calculation of the effect of such orbit errors on the determination of the
relativistic frequency shift of the clocks and on the time transfer (MWL) for the ACES mission (section \ref{results}). We provide the overall
requirements on orbit determination for the ACES mission, and show that the mission objectives can be achieved with relatively modest performance on
orbit determination. We discuss those results and conclude in section \ref{conclusion}.

\section{The ACES mission}
\label{ACES}

The ACES project led by CNES and ESA aims at setting up on the ISS several highly stable clocks in 2013. The ACES payload includes two clocks, a
hydrogen maser (SHM developed by TEMEX) and a cold atom clock PHARAO (developed by CNES) respectively for short and long term performances, and a
microwave link for communication and time/frequency comparison. The frequency stability of PHARAO onboard the ISS is expected to be better than
$10^{-13}$ for one second, $3 \cdot 10^{-16}$ over one day and $1 \cdot 10^{-16}$ over ten days, with an accuracy goal of $1 \cdot 10^{-16}$ in
relative frequency.

The ACES mission has as objectives :
\begin{itemize}
    \item to operate a cold atom clock in microgravity with a 100 mHz
    linewidth,
    \item to compare the high performances of the two atomic clocks in space
    (PHARAO and SHM) and to obtain a stability of $3\cdot 10^{-16}$
    over one day,
    \item to perform time comparisons between the two space clocks and ground clocks,
    \item to carry out tests of fundamental physics such as a gravitational redshift
    measurement and a test of Lorentz invariance, and to search for a possible drift of the fine structure constant $\alpha$.
    \item to perform precise measurements of the Total Electron Content (TEC) in the ionosphere, the
    tropospheric delay and the Newtonian potential.
 \end{itemize}

The time transfer is performed using a micro-wave two-way system,
called Micro-Wave Link (MWL). An additional frequency is added in
order to measure and correct the ionospheric delay at the required
level. It uses carriers of frequency 13.5, 14.7 et 2.25 GHz,
modulated by pseudo random codes respectively at $10^{8}$
$s^{-1}$, $10^{8}$ $s^{-1}$ and $10^{6}$ $s^{-1}$ chip rates.
Moreover it has four channels that allow four ground stations to
be compared with the ISS clock at the same time.

According to the mission specifications, the microwave link has to
synchronize two atomic clocks with a time stability of
$\leq$~0.3~ps over 300~s, $\leq$~7~ps over one day, and
$\leq$~23~ps over 10 days. The performance of this link is a key
issue since it will perform high precision time comparisons
without damaging the high performances of the clocks.

For our purposes we express the above requirements for the MWL in a simplified form by the temporal Allan deviation $(\sigma_x(\tau))$ expressed in
seconds:

\begin{equation}
\sigma_x (\tau) = 5.2 \cdot 10^{-12} {\rm s} \cdot \tau^{-\frac{1}{2}} \label{BruitIntrinseque}
\end{equation}
with the integration time $\tau$ in seconds. Equation (\ref{BruitIntrinseque}) is valid for a single satellite pass over a ground station (for
integration times $\tau$ lower than $300$ s). For longer integrations times

\begin{equation}\label{AttentesAuxTempsLongs}
\sigma_x (\tau) = 2.4 \cdot 10^{-14}{\rm s} \cdot \tau^{\frac{1}{2}}.
\end{equation}


The temporal Allan deviation can be related to frequency
instability as expressed by the modified Allan deviation $Mod
\sigma_y(\tau)$ by

\begin{equation}\label{modsigmay}
Mod\sigma_y(\tau) = \frac{\sqrt{3}\ \sigma_x (\tau)}{\tau}.
\end{equation}

We take (\ref{BruitIntrinseque}) and (\ref{AttentesAuxTempsLongs}) as our upper limits for the calculation of all perturbing effects in the following
sections, together with the overall accuracy requirement (maximum allowed frequency bias) of $1 \cdot 10^{-16}$ in relative frequency.

\section{Relativistic model for clocks and time transfer of ACES}
\label{relat}

In a general relativistic framework each clock produces its own
local proper time, in our case $\tau^g$ and $\tau^s$ for the
ground and space clocks respectively.

In order to model signal propagation between the ground and the space stations, we use a non-rotating geocentric space-time coordinate system. Thus
$t=x_0/c$ is the geocentric coordinate time, $\overrightarrow{x} = (x_1, x_2, x_3)$ are the spatial coordinates, where c is the speed of light in
vacuum  ($c = 299792458$~m.s$^{-1}$). We denote $U(t,\overrightarrow{x})$ as the total Newtonian potential at the coordinate time $t$ and the
position $\overrightarrow{x}$ with the convention that $U \geq 0$ \cite{Soffel2}. In these coordinates, the metric is given by an approximate
solution of Einstein's equations  valid for low velocity and potential ($\frac{U}{c^{2}}<<1$ and $\frac{v^{2}}{c^{2}}<<1$):

\begin{equation}
ds^{2} = -(1 - \frac{2 U(t, \overrightarrow{x})}{c^2})c^2 dt^2 +(1
+ \frac{2 U(t, \overrightarrow{x})}{c^2})d\overrightarrow{x}^2,
\end{equation}
where higher order terms can be neglected for our purposes
\cite{Wolf1}.

\begin{figure}
        \includegraphics[height=3.6cm]{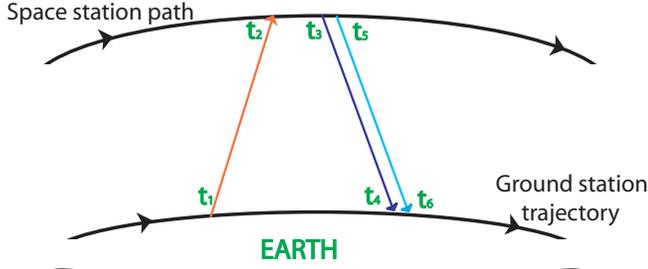}
    \caption{MWL principle}
    \label{fig:ACESTrajectoires}
\end{figure}

In this system, each emission or reception event (at the antenna
phase center) is identified by its coordinate time $t_i$ (figure
\ref{fig:ACESTrajectoires}) and a coordinate time interval is
defined by $T_{ij}=t_{j}-t_{i}$. We define $\overrightarrow{x}_g$,
$\overrightarrow{v}_g$ and $\overrightarrow{a}_g$ respectively as
the position, the velocity and the acceleration of the ground
station, and $\overrightarrow{x}_s$, $\overrightarrow{v}_s$ and
$\overrightarrow{a}_s$ respectively as the position, the velocity
and the acceleration of the space station.

The ACES mission uses two different antennas: one Ku-band antenna
for uplink and downlink, and one S-band antenna for downwards
signal only. The antennae are characterized by their phase
pattern, which describes the position of the antenna phase center
as a function of the direction of the incoming signal. The phase
patterns of the MWL antennas have been measured in the laboratory
at all three frequencies. For example, the phase variations with
direction of the Ku-band antenna can reach up to $\sim$0.1~rad
($\sim$1~ps in time). Those variations can be corrected from the
known phase pattern and a knowledge of ISS attitude (ie. direction
of the incoming signal). Once corrected the remaining errors,
mostly due to uncertainties in ISS attitude, are below the MWL
specifications.

The $f_1$ frequency signal is emitted by the ground station at the
coordinate time $t_1$ and received by the space station at $t_2$.
The $f_2$ and $f_3$ frequency signals are emitted from the space
station at $t_3$ and $t_5$, and received at the ground station at
$t_4$ and $t_6$. The third frequency is added to measure the TEC
in the ionosphere which allows the correction of the ionospheric
delay.

The MWL is characterized by its continuous way of emission. It
measures the time offsets between the locally generated signal and
the received one. It provides three measurements (or observables)
of the code (one on the space station, two on the ground) and
three measurements of the phase of the carrier frequency at a
sampling rate of one Hertz.

An observable is related to the phase comparison between a signal derived from the local oscillator and the received signal, corrected for the
frequency difference mainly due to the first order Doppler effect (see Bahder T. B. 2003 for details of a similar procedure used in GPS). If we
consider a particular bit of the signal which is produced locally at $\tau_p$ and received at $\tau_a$, an observable is a measurement of the local
proper time interval between these two events. The actual measurement of the phase difference $\Delta\Phi (\tau_a, \overrightarrow{x}_a)$ occurs at
the single space-time point ($\tau_a$ ,$\overrightarrow{x}_a$) and is labeled with the arrival proper time $\tau_a$. It can be expressed as

\begin{equation}
\Delta \tau (\tau_a, \overrightarrow{x}_a)=  \frac{\Delta\Phi (\tau_a, \overrightarrow{x}_a)}{\omega} + \delta\tau = \tau_p - \tau_a
\end{equation}
where $\omega$ is a conversion factor from phase to proper time depending on the nominal carrier and code frequencies, and $\delta\tau$ represents
the measurement errors (difference between the clock and ideal proper time, measurement phase noise, etc...).

%
%
%

Considering the experimental uncertainties of the ACES mission
(see equation (\ref{BruitIntrinseque}) and
(\ref{AttentesAuxTempsLongs})), we will neglect any terms in the
relativistic model that, when maximised, lead to corrections of
less than $3 \cdot 10^{-13}$ s in time. Numerical applications are
necessary to evaluate which terms can be neglected. For this
purpose, we consider the International Space Station with a
circular trajectory at an altitude of 400~km in a plane inclined
by $51.6^{o}$ with respect to the equatorial plane. It has a
velocity $v_s = 7.7\cdot 10^3$ $m.s^{-1}$ in a gravitational
potential of $U_s/c^2 = 6.5 \cdot 10^{-10}$. The ground station
has a velocity $v_g = 465$ $m.s^{-1}$ at a gravitational potential
of $U_g/c^2 = 6.9 \cdot 10^{-10}$.

The tropospheric delay $\Delta^{tropo}$ is considered as independent of the frequency of the signal, with a slow variation with time and an amplitude
of less than 100~ns.

We assume the density of electrons in the ionosphere $N_e$ is less
than $2 \cdot 10^{12}$ electrons/$m^{3}$. The ionospheric delay
$\Delta^{iono}$ is maximum for the $f_3$
 frequency signal with an amplitude of less than 10~ns.

The relation between the proper time $\tau$ and the coordinate
time $t$ is given to sufficient accuracy by

\begin{equation}
\frac{d \tau}{d t} = 1 - \left(
\frac{U(t,\overrightarrow{x})}{c^{2}} + \frac{v^{2}(t)}{2 c^{2}}
\right) + O(c^{-4}). \label{Blanchet}
\end{equation}

Higher order terms of equation (\ref{Blanchet}) have negligible
effects at the projected uncertainty of $1 \cdot 10^{-16}$ in
relative frequency of the ACES clocks \cite{Wolf1}. Note however,
that some care has to be taken when evaluating the Newtonian
potential $U(t,\overrightarrow{x})$ in (\ref{Blanchet}) for the
ground or space clock \cite{Wolf1}.

The ACES mission aims at obtaining the variation of the desynchronisation between ground and space clocks with time, that is to say, the function
$\tau^g(t) - \tau^s(t)$. It is evaluated by combining the measurements performed on the ground and onboard the space station and a precise
calculation of the signal propagation times. In order to be able to control $T_{23}$ (see below), we combine two measurements
$\Delta\tau^s\left(\tau^s(t_2)\right)$ and $\Delta \tau^g\left(\tau^g(t_4)\right)$ but with $\tau^s(t_2) \neq \tau^g(t_4)$. Then the expression of
desynchronisation reads (see \cite{DuchaynePhD} for a detailed derivation)

\begin{equation}
\begin{split}
\tau^g(t_a) - \tau^s(t_a)
 &=\frac{1}{2}\biggl(\Delta\tau^s\left(\tau^s(t_2)\right) - \Delta
\tau^g\left(\tau^g(t_4)\right)\\ &+ T_{12} - T_{34}\\
&- \int_{t_1}^{t_2}(\frac{U (t, \overrightarrow{x_g})}{c^2} +
\frac{v_g^{2}(t)}{2 c^2})dt \\
& + \int_{t_3} ^{t_4}(\frac{U (t,
\overrightarrow{x_s})}{c^2} + \frac{v_s^{2}(t)}{2 c^2})dt \biggr),\\
\label{ExpressionDesynMWLReceptionTime}
\end{split}
\end{equation}

where $t_a = \frac{t_2 + t_4}{2}$, and where
$\Delta\tau^s\left(\tau^s(t_2)\right)$ and $\Delta
\tau^g\left(\tau^g(t_4)\right)$ are the observables respectively
from the ground and onboard the satellite at the coordinate times
$t_2$ and $t_4$, and where we have neglected non-linearities of
$\tau^g(t)$ and $\tau^s(t)$ over the interval $t_4 - t_2$ (few
milliseconds). The integral terms result from proper time to
coordinate time transformations. They are small corrections of
order $10^{-12}$ s to the desynchronisation $\tau^g(t_a) -
\tau^s(t_a)$.

However, it is the derivative with respect to the coordinate time
$t$ of the relation (\ref{ExpressionDesynMWLReceptionTime}) which
has to be studied for applications such as the gravitational
redshift test or geodesy. Actually it has to be compared with the
next relation obtained from equation (\ref{Blanchet}):

\begin{equation}
\begin{split} \frac{d \tau^g}{d t}- \frac{d \tau^s}{d t} &=
 \frac{1}{c^{2}}\cdot \biggl(U(t,\overrightarrow{x_s}) - U(t,\overrightarrow{x_g})\\ &+ \frac{v_s^{2}(t)}{2} - \frac{v_g^{2}(t)}{2}\biggr)+
 O(c^{-4}).\\
\end{split}
\label{ExpressionRedshift}
\end{equation}

We note that any constant term appearing in the desynchronisation
expression (\ref{ExpressionDesynMWLReceptionTime}) will have no
effect on the final result (\ref{ExpressionRedshift}) because of
the derivation.

In (\ref{ExpressionDesynMWLReceptionTime}) the difference
$T_{12}-T_{34}$ needs to be calculated from the knowledge of
satellite and ground positions and velocities (orbit
determination). For example, $T_{12}$ is the time interval elapsed
between emission from the ground station and reception by the
satellite of the $f_1$ frequency signal. It can be written as

\begin{equation}
\begin{split}
T_{12} &= \frac{R_{12}}{c} + \frac{2 G
M_{E}}{c^{3}}\ln\left(\frac{x_{g}(t_1) + x_{s}(t_2) +
R_{12}}{x_{g}(t_1) +x_{s}(t_2) - R_{12}}\right)\\& +
\Delta_{12}^{tropo}+ \Delta_{12}^{iono}+O(\frac{1}{c^4}), \\
\label{ExpressionT12}
\end{split}
\end{equation}
where $R_{12} = ||\overrightarrow{R_{12}}|| =
||\overrightarrow{x_s}(t_2)-\overrightarrow{x_g}(t_1)||$, where
the logarithmic term represents the Shapiro time delay
\cite{Shapiro} (see e.g. Blanchet L. et al. 2001 for a detailed
derivation) and where $\Delta_{12}^{tropo}$ and
$\Delta_{12}^{iono}$ are respectively the tropospheric and
ionospheric delays on the signal path.

Only one of the clocks (here we assume the ground clock) has a
known relation with coordinate time $t$, so
$\overrightarrow{x}_s(t_2)$ cannot be directly obtained from the
orbit determination, only $\overrightarrow{x}_s(t_1)$ is known.
Relation (\ref{ExpressionT12}) is then modified to

\begin{equation}
\begin{split}
T_{12} &= \frac{D(t_1)}{c} +
\frac{\overrightarrow{D}(t_1).\overrightarrow{v_{s}}(t_1)}{c^{2}}
\\
&+ \frac{D(t_1)}{2 c^{3}}\biggl(||\overrightarrow{v_{s}}(t_{1})
||^{2} + \overrightarrow{D}(t_1).\overrightarrow{a_{s}}(t_1)\\
 &+ (\frac{\overrightarrow{D}(t_1).\overrightarrow{v_{s}}(t_1)
}{D(t_1)})^{2}\biggr)\\
& + \frac{2 G M_{E}}{c^{3}}\ln\left(\frac{x_{g}(t_1) + x_{s}(t_1)
+ D(t_1)}{x_{g}(t_1) +x_{s}(t_1) - D(t_1)}\right)\\&
+\Delta_{12}^{iono}+ \Delta_{12}^{tropo} +O(c^{-4}),\\
\end{split}
\label{ExpressionT12_D}
\end{equation}
where $\overrightarrow{D}(t) =
\overrightarrow{x_s}(t)-\overrightarrow{x_g}(t)$, $D(t) =
||\overrightarrow{D}(t)||$, and where the unknown position
$\overrightarrow{x_s}(t_2)$ in (\ref{ExpressionT12}) was expanded
in a Taylor series around the corresponding known position
$\overrightarrow{x_s}(t_1)$. The ionospheric and tropospheric
terms are related to the signal paths. They need to be calculated
for $\overrightarrow{R}(t)$, but only $\overrightarrow{D}(t)$ is
known. The corresponding correction is related to the motion of
the space station during the atmospheric delay of the signal (the
atmospheric equivalent of the "Sagnac" type term ($2^{nd}$ term in
(\ref{ExpressionT12_D})). The corresponding corrections to the
tropospheric term and to the ionospheric terms at frequencies
$f_1$ and $f_2$ are negligible but included for the $f_3$ signal
(see equation (\ref{ExpressionTEC})).

Assuming that $T_{23} \leq 10^{-3}$ s , the expression of
$T_{12}-T_{34}$ where all terms greater than $3 \cdot 10^{-13}$~s
have been included when maximized for ACES can be evaluated by~:

\begin{equation}
\begin{split}
T_{12} - T_{34} &=  2
\frac{\overrightarrow{D}(t_4).\overrightarrow{v_g}(t_4)}{c^2} +
\frac{\overrightarrow{D}(t_4).\overrightarrow{\Delta v}(t_4)}{c
\cdot D(t_4)} T_{23} \\
&+ 2 \frac{D(t_4)}{c^3}\cdot \biggl(\overrightarrow{\Delta
v}(t_4).\overrightarrow{v_s}(t_4)\\& -
\overrightarrow{D}(t_4).\overrightarrow{a_g}(t_4) + \|
\overrightarrow{\Delta v}(t_4) \|^2 \biggr)\\&+
\frac{T_{23}}{c^2}\cdot \biggl(\overrightarrow{\Delta
v}(t_4).\overrightarrow{v_s}(t_4)\\&
 - \overrightarrow{D}(t_4).\overrightarrow{a_s}(t_4) + 2\| \overrightarrow{\Delta v}(t_4) \|^2 \\
&- 2\overrightarrow{D}(t_4).\overrightarrow{\Delta a}(t_4) - (\frac{\overrightarrow{D}(t_4).\overrightarrow{\Delta v}(t_4)}{D(t_4)})^2 \biggr)\\
&+ \frac{T_{23}^2}{2 c D(t_4)}\biggl( \| \overrightarrow{\Delta v}(t_4) \|^2 - \overrightarrow{D}(t_4).\overrightarrow{\Delta a}(t_4) \\
&- (\frac{\overrightarrow{D}(t_4).\overrightarrow{\Delta
v}(t_4)}{D(t_4)})^2  \biggr)\\&+\Delta_{12}^{iono} -
\Delta_{34}^{iono}
+O(\frac{1}{c^4}),\\
\label{ExpressionT12T34}
\end{split}
\end{equation}
where $\overrightarrow{\Delta
v}(t)=\overrightarrow{v}_g(t)-\overrightarrow{v}_s(t)$ and
$\overrightarrow{\Delta
a}(t)=\overrightarrow{a}_g(t)-\overrightarrow{a}_s(t)$.

To obtain (\ref{ExpressionT12T34}) we have applied
(\ref{ExpressionT12_D}) for the upward and downward signals and
expanded all positions and velocities in Taylor series around
their values at $t_4$, which can be obtained from the time of
measurement ("label" of the observable $\Delta\tau^g(\tau^g(t_4))$
in (\ref{ExpressionDesynMWLReceptionTime})), on the ground.

The difference $T_{12} - T_{34}$ of upward and downward signals at
$f_1$ and $f_2$ allows to eliminate to first order delaying and
restraining factors such as range $(D/c)$, troposphere or Shapiro
effects. Due to the asymmetry of the paths, that cancellation is
not perfect, and there are some terms left (equation
(\ref{ExpressionT12T34})) which depend on orbit determination as
well as on the coordinate time interval $T_{23}$ elapsed between
reception and emission at the phase centre of the MWL antenna
onboard the ISS.

The time interval $T_{23}$ can be controlled by "shifting" the
time of the measurements on the ground or on board the space
station (e.g. shifting $\tau^s(t_2)$ in
$\Delta\tau^s(\tau^s(t_2))$ of equation
(\ref{ExpressionDesynMWLReceptionTime})) when post-processing the
data. That allows the control of the difference between emission
and reception {\it at the clock}. But $T_{23}$ in
(\ref{ExpressionT12T34}) is defined at the antenna phase centre,
therefore its control requires the knowledge (calibration and in
situ measurement) of the instrumental delays (cables, electronics,
etc...) in the MWL space segment. So the control and knowledge
(uncertainty $\delta T_{23}$) of $T_{23}$ is determined by the
uncertainty of the calibration and measurement of internal delays,
which will play a role in the following.

The ionospheric effect $\Delta^{iono}$ on a radio signal of frequency $f$ can be modelled as follows \cite{Bassiri}:

\begin{equation}
\begin{split} \Delta^{iono} &=\frac{40.3082}{c \cdot f^{2}} \int N_e dL\\&
- \frac{7527}{f^{3}} \int
(\overrightarrow{B}_0.\overrightarrow{k})\cdot N_e dL
+O(\frac{1}{c^4}),
\end{split}
 \label{Kedar}
\end{equation}

where $\overrightarrow{B}_0$ is the Earth's magnetic field
($||\overrightarrow{B}_0|| \approx 3.12 \cdot 10^{-5} T$) and
$\overrightarrow{k}$ is the direction of propagation of the
signal.

The electron density integrated along the signal trajectory, $\int
N_e dL$, is usually referred as the Total Electron Content (TEC).
The difference of ionospheric delays appearing in equation
(\ref{ExpressionT12T34}) is then

\begin{equation}
\begin{split}
\Delta_{12}^{iono} - \Delta_{34}^{iono} &= (\frac{1}{f_{1}^{2}} -
\frac{1}{f_{2}^{2}})\frac{40.3082}{c} TEC \\
&- \frac{7527}{f_1^{3}}\int
\frac{\overrightarrow{B}_0.\overrightarrow{D}(t_2)}{D(t_2)}N_e
dL\\
 &- \frac{7527}{f_2^{3}} \int\frac{
\overrightarrow{B}_0.\overrightarrow{D}(t_4)}{D(t_4)}N_e dL+
O(\frac{1}{c^4}).\\
\end{split}
\label{iono}
\end{equation}

The second and third term in (\ref{iono}) represent the effect of
the Earth's magnetic field and amount to at most 0.5 ps for ACES,
which is almost negligible. Therefore, an only very rough
estimation of these terms is sufficient.

The TEC is determined by combining the two downwards signal observables. The time interval $T_{46}$ elapsed between the receptions of the two signals
depends on the differences of their internal delays at the emission from the space station (i.e. $T_{35}$) and on the difference of their propagation
time. Assuming that $T_{46}< 1~\mu s$, the expression of the Total Electron Content is then given by the following expression :

\begin{equation}
\begin{split}
TEC &= \frac{c}{40.3082}\frac{1}{\frac{f_{2}^{2} -
f_{3}^{2}}{f_{2}^{2} f_{3}^{2}}+\frac{f_{2}^{3} -
f_{3}^{3}}{f_{2}^{3} f_{3}^{3}} \frac{7527 \cdot c}{40.3}
\frac{\overrightarrow{B}_0.\overrightarrow{D}(t_4)}{D(t_4)}}\times\\
&\biggl( \Delta \tau^g\left(\tau^g(t_4)\right)  - \Delta
\tau^g\left(\tau^g(t_6)\right) \\&+
\frac{\overrightarrow{D}(t_4).\Delta\overrightarrow{v}(t_4)}{D(t_4)}
\frac{T_{46}}{c} \biggr) / \biggl(1-\\&
\frac{\overrightarrow{D}(t_4).\overrightarrow{v_{s}}(t_4) }{D(t_4)
\cdot c}\biggr)+ O(\frac{1}{c^4}). \\
\label{ExpressionTEC}
\end{split}
\end{equation}

The last term in (\ref{ExpressionTEC}) is negligible when used for
the time transfer (when inserted into (\ref{iono})) but may be
significant for the study of the TEC itself.

In summary, a reliable orbit determination is required for two
main reasons. On one hand to calculate precisely the corrections
in equations (\ref{ExpressionT12T34}). On the other hand, to
evaluate correctly the terms on the right hand side of equation
(\ref{ExpressionRedshift}) ie. the second order Doppler and
gravitational redshifts.

In addition, we also need a precise knowledge of the time interval
$T_{23}$, (ie. of the onboard internal delays) in order to be able
to calculate the corresponding terms in (\ref{ExpressionT12T34})
with sufficient accuracy.

The aim of this work is to estimate using simple orbit error models, which levels of uncertainty on orbit determination and calibration of internal
delays (knowledge of $T_{23}$) are required to reach the required performances. For that purpose only the leading terms in (\ref{ExpressionT12T34})
are required ie.

\begin{equation}
\begin{split} T_{12} - T_{34} &=  2
\frac{\overrightarrow{D}(t_4).\overrightarrow{v_g}(t_4)}{c^2}\\ &+
\frac{\overrightarrow{D}(t_4).\overrightarrow{\Delta v}(t_4)}{c
\cdot D(t_4)} T_{23}+O(\frac{1}{c^3}),\\
\end{split}\label{T12T34court}
\end{equation}
which, together with equation (\ref{ExpressionRedshift}) for the relativistic frequency shift, is sufficient to derive the maximum allowed
uncertainties on orbit determination and internal delays in order to stay below the limits given by (\ref{BruitIntrinseque}) and
(\ref{AttentesAuxTempsLongs}).

\section{Effects of orbit errors on clock comparison}
\label{Effects}

We now use equations (\ref{T12T34court}) and (\ref{ExpressionRedshift}) to express the effects of station trajectory and time calibration
uncertainties on the time transfer and on the relativistic frequency shift. In this section we make no assumptions on orbit determination errors, our
results being completely general and valid up to the neglected terms as described below.

We note ($\overrightarrow{X}_a(t)$, $\overrightarrow{V}_a(t)$) and ($\overrightarrow{X}_a'(t)$, $\overrightarrow{V}_a'(t)$) respectively the true and
computed trajectories of the antenna phase center, and ($\overrightarrow{X}_c(t)$, $\overrightarrow{V}_c(t)$) and ($\overrightarrow{X}_c'(t)$,
$\overrightarrow{V}_c'(t)$) respectively the true and computed trajectories of the clock reference point. We also define ($\overrightarrow{X}_o(t)$,
$\overrightarrow{V}_o(t)$) the true trajectory and the true velocity of the center of mass of the ISS. These five trajectories are expressed in the
non-rotating geocentric frame (GCRS, Geocentric Celestial Reference System) \cite{Soffel2}.

On one hand, the error in the time transfer is related with the uncertainties of $T_{12} - T_{34}$, and can be obtained from the simplified equation
(\ref{T12T34court}). It is then dependent on the ground and space station trajectory knowledge, on the value of $T_{23}$ and on the uncertainty on
this parameter $\delta T_{23}$. As said before, a precise knowledge of the time interval $T_{23}$ is related to the internal delay calibrations. The
error on $T_{12} - T_{34}$ is

\begin{equation}
\begin{split}
\delta \left(T_{12} - T_{34}\right) &=  2 \frac{\delta \overrightarrow{D}.\overrightarrow{v_g}+\overrightarrow{D}.\delta\overrightarrow{v_g}}{c^2}
+\frac{\overrightarrow{D}.\Delta\overrightarrow{ v}}{c \cdot D}
\delta T_{23}\\
&+ \biggl(\frac{\delta \overrightarrow{D}.\Delta\overrightarrow{ v}}{c \cdot D} + \frac{\overrightarrow{D}.\delta\Delta\overrightarrow{v}}{c \cdot D}
\\&- \frac{\overrightarrow{D}.\Delta\overrightarrow{ v}}{c \cdot D}\frac{\delta D}{D} \biggr) T_{23}.
\end{split}
\end{equation}

In general ground station uncertainties are below 10 cm. Thus the uncertainty on ground station position is negligible with respect to the ISS
position errors and the knowledge of the vector $\overrightarrow{D}$ is only related to the uncertainty of the space station reference point which is
the antenna phase center. Then we have
 $\delta \overrightarrow{D} = \overrightarrow{X_a X_a'}$.

The previous equation can then be written as :

\begin{equation}
\begin{split}
 \delta \left(T_{12} - T_{34}\right) &=  2
\frac{\overrightarrow{X_a X_a'}.\overrightarrow{v_g}}{c^2} +\frac{\overrightarrow{D}.\overrightarrow{\Delta v}}{c \cdot D}
\delta T_{23}\\
&+ \biggl(\frac{\overrightarrow{X_a X_a'}.\overrightarrow{\Delta v}}{c \cdot D} - \frac{\overrightarrow{D}}{c \cdot D}.\frac{d\overrightarrow{X_a
X_a'}}{dt} \\&- \frac{\overrightarrow{D}.\overrightarrow{\Delta v}}{c \cdot D}\frac{\|\overrightarrow{X_a X_a'}\|}{D} \biggr) T_{23}.
 \label{FinalEquationTimeTransfer}
\end{split}
\end{equation}

We note that equation (\ref{FinalEquationTimeTransfer}) depends the antenna phase center position error ($\overrightarrow{X_a X_a'}$), the coordinate
time interval $T_{23}$ and the uncertainty on this time interval $\delta T_{23}$.

On the other hand, the clock relativistic correction along a trajectory is obtained from equation (\ref{ExpressionRedshift}). It depends on the
position and the velocity of the reference point, in this case the clock. We need to express the error on the reference point frequency - ie. the
frequency difference between the true clock position and the computed clock position - in order to compare its Modified Allan deviation with the
specifications.

The gravitational potential can be evaluated on a given trajectory with sufficient precision \cite{Wolf1} using gravity models (eg. GRIM5 or EGM96).
The error on the frequency shift at the clock position is given by~:

\begin{equation}
\begin{split} \delta(\frac{d \tau}{d t})_{\overrightarrow{X_c}} =& (\frac{d
\tau}{d t})_{\overrightarrow{X_c}} - (\frac{d \tau}{d t})_{\overrightarrow{X_c'}}\\=& - \frac{1}{c^{2}}\biggl( U(t,\overrightarrow{X_c}) -
U(t,\overrightarrow{X_c'})\\& + \frac{V_c^2 - V_c'^2}{2}\biggr). \label{ExpressionRedchift}
\end{split}
\end{equation}

The frequency difference between the reference point position $\overrightarrow{X}$ and the ISS center of mass $\overrightarrow{X}_o$ is given by :

\begin{equation}
\begin{split} (\frac{d \tau}{d t})_{\overrightarrow{X}} - (\frac{d
\tau}{d t})_{\overrightarrow{X}_o}=& - \frac{1}{c^{2}}\biggl( U(t,\overrightarrow{X}) - U(t,\overrightarrow{X}_o) \\&+ \frac{V^2 - V_o^2}{2}\biggr).
\label{ExpressionRedchift}
\end{split}
\end{equation}

The trajectory of $\overrightarrow{X}_o$ is the solution of the differential equation

\begin{equation}\frac{d^2 \overrightarrow{X}_o}{d t^2} = \overrightarrow{\Gamma}_P +
\overrightarrow{\Gamma}_S, \label{PdfISSCenterOfMass}
\end{equation}

where $\overrightarrow{\Gamma}_P$ is the acceleration due to the gravitational potential and $\overrightarrow{\Gamma}_S$ is the acceleration due to
other effects (e.g. surface forces like air drag and radiation pressure).

Using  \begin{equation} \overrightarrow{\Gamma}_P =\overrightarrow{Grad} (U),\end{equation}

and to first order

\begin{equation} U(t,\overrightarrow{X}) =
U(t,\overrightarrow{X}_o) + \overrightarrow{\Gamma}_P(\overrightarrow{X_o}). \overrightarrow{X_o X},\end{equation}

and multiplying equation (\ref{PdfISSCenterOfMass}) by the vector $\overrightarrow{X_o X}$ (which is the position of the clock with respect to the
ISS center of mass), we can substitute the difference of gravitational potential in equation (\ref{ExpressionRedchift}) by :

\begin{equation}
\begin{split}U(t,\overrightarrow{X}) -
U(t,\overrightarrow{X}_o) &= \overrightarrow{\Gamma}_P(\overrightarrow{X_o}). \overrightarrow{X_o X} \\&= \frac{d^2 \overrightarrow{X}_o}{d
t^2}.\overrightarrow{X_o X} - \overrightarrow{\Gamma}_S.\overrightarrow{X_o X} .
\end{split}\end{equation}

Then we obtain :

\begin{equation}
\begin{split}(\frac{d \tau}{d t})_{\overrightarrow{X}} - (\frac{d \tau}{d t})_{\overrightarrow{X}_o}
&=- \frac{1}{c^{2}} \biggl( \frac{d \overrightarrow{V}_o} {d t}.\overrightarrow{X_o X}+ \overrightarrow{V_o}.\frac{d \overrightarrow{X_o X}}{dt} \\&+
\frac{1}{2} (\frac{d \overrightarrow{X_o X}}{dt})^2- \overrightarrow{\Gamma}_S.\overrightarrow{X_o X} \biggr), \label{ExpressionRedchift2}
\end{split}
\end{equation}


which can be simplified to :

%

\begin{equation}
\begin{split}(\frac{d \tau}{d t})_{\overrightarrow{X}} -
(\frac{d \tau}{d t})_{\overrightarrow{X}_o}& =- \frac{1}{c^{2}} \biggl( \frac{d } {d t}\left(\overrightarrow{V_o}.\overrightarrow{X_o X} \right)\\&+
\frac{1}{2} (\frac{d \overrightarrow{X_o X}}{dt})^2- \overrightarrow{\Gamma}_S.\overrightarrow{X_o X} \biggr).
\end{split}
\end{equation}

In this expression the term $\frac{1}{2} (\frac{d \overrightarrow{X_o X}}{dt})^2- \overrightarrow{\Gamma}_S.\overrightarrow{X_o X}$ can be
interpreted as the relativistic correction for the clock referenced to the local ISS frame. In fact, for an observer in the ISS frame, the non
gravitational acceleration $\overrightarrow{\Gamma}_S$ produces an acceleration which can be seen by this observer as coming from a "gravitational
potential" $-\overrightarrow{\Gamma}_S.\overrightarrow{X_o X}$. The term $\frac{d } {d t}\left(\overrightarrow{V_o}.\overrightarrow{X_o X} \right)$
is the position of the clock with respect to the ISS center of mass projected on the ISS velocity direction (along track component). If there are no
external accelerations, and if the velocity of $\overrightarrow{X}$ in the ISS frame remains small, only this term is present.

Combining the previous equation with the same expressed in $\overrightarrow{X'}$, we obtain :

\begin{equation}
\begin{split} \delta(\frac{d \tau}{d t})_{\overrightarrow{X_c}} &= \frac{1}{c^{2}}
\biggl( \frac{d } {d t}\left(\overrightarrow{V_o}.\overrightarrow{X_c X_c'}  \right)+ \frac{1}{2} (\frac{d \overrightarrow{X_o
X_c}}{dt})^2\\
&- \frac{1}{2} (\frac{d \overrightarrow{X_o X_c'}}{dt})^2 - \overrightarrow{\Gamma}_S.\overrightarrow{X_c X_c'} \biggr). \label{ExpressionRedchift3}
\end{split}
\end{equation}

In order to simplify equation (\ref{ExpressionRedchift3}), we evaluate the order of magnitude of the different contributors appearing in this
equation.

\begin{figure}[h!]
\includegraphics[height=5.5cm]{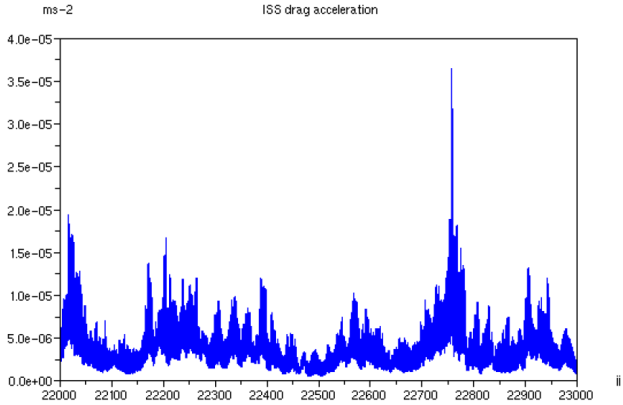}
    \caption{Estimation of the non gravitational acceleration of the ISS vs time (in days)}
    \label{ISSDrag}
\end{figure}

To investigate the importance of the non gravitational term $\frac{\overrightarrow{\Gamma}_S.\overrightarrow{X_c X_c'}}{c^2}$, the drag has been
modeled along a reference orbit of the ISS. A period with important solar activity has been chosen in order to evaluate the worst case (see figure
\ref{ISSDrag}).

 To estimate its effect
on formula (\ref{ExpressionRedchift3}), the acceleration has been multiplied by a 10 meter bias, by a 10 meter random noise or by a 10 meter
sinusoidal function at orbital period, corresponding to possible attitude and orbit error effects of the ISS. The Allan deviation stays below
$10^{-21}$, which is totally negligible here. Also, this term has no effect on the frequency accuracy at the $10^{-16}$ level aimed at by ACES. In
addition, the residual term of the second order Doppler shift $\frac{1}{2 c^2} \left[(\frac{d \overrightarrow{X_o X_c}}{dt})^2-(\frac{d
\overrightarrow{X_o X_c'}}{dt})^2 \right]$ must be computed with the GCRS trajectories. The order of magnitude of these terms can be evaluated as
$\sim (a \delta a \Omega^2/{c^2})$ with $\Omega$ the orbital angular frequency, $a = ||\overrightarrow{X_o X_c}||$, and $\delta a$ its error. For $a
\leq 100$~m and $\delta a \leq 10$~m this effect is also totally negligible.

The only important term for the performance evaluation is thus the along track term  $\frac{1}{c^{2}} \frac{d } {d
t}\left(\overrightarrow{V_o}.\overrightarrow{X_c X_c'}\right)$, and equation (\ref{ExpressionRedchift3}) can be written :

\begin{equation}
\begin{split} \delta(\frac{d \tau}{d t})_{\overrightarrow{X_c}}& = (\frac{d
\tau}{d t})_{\overrightarrow{X_c}} - (\frac{d \tau}{d t})_{\overrightarrow{X_c}'} \\&= \frac{1}{c^{2}} \biggl[ \frac{d } {d
t}\left(\overrightarrow{V_o}.\overrightarrow{X_c X_c'} \right) \biggr]. \label{RedshiftFinal}
\end{split}
\end{equation}

So only the component of the clock position error $\overrightarrow{X_c X_c'}$ projected on the velocity of the ISS $\overrightarrow{V_o}$ plays a
role. This can be understood considering for example a purely positive radial component. In this case we underestimate the gravitational potential
but overestimate the velocity, and the two cancel. We underline again the fact that the derivations in this chapter are valid for any type of orbital
errors, up to the neglected terms described above.

The scalar products of vectors can be evaluated in a local frame : for example it may be useful to study them in the local orbital frame
($\overrightarrow{R}$, $\overrightarrow{T}$, $\overrightarrow{N}$) defined with $\overrightarrow{R}$ the unit vector between the Earth's center and
the space station, $\overrightarrow{N}$ orthogonal to $\overrightarrow{R}$ and the inertial velocity, and $\overrightarrow{T}$ orthogonal to
$\overrightarrow{R}$ and $\overrightarrow{N}$.

The ISS attitude is assumed to be roughly constant. The clock position error in this frame is represented by a constant vector (error in the position
of the clock relative to the centre of mass), and by perturbations which reflect :
\begin{itemize}
\item  attitude uncertainties (rigid body behaviour of ISS).
If we take $\pm 5$ degrees uncertainty and a distance of 10 meters, this leads to 0.87 meters amplitude error for the clock position.

\item  ISS deformations : they are due to thermoelastic effects
which will be mainly at orbital period (eclipses and sun orientation). We suppose that these effects are below one meter amplitude.

\item  vibrations : here we suppose that the 'vibrations' are
related to frequencies higher or equal to the first ISS eigenfrequency which is higher than the orbital period. The corresponding displacements
remain small and are expected to stay below one meter.
\end{itemize}
Either of these effects is negligible when inserted into (\ref{FinalEquationTimeTransfer}) and (\ref{RedshiftFinal}). So for our purposes the only
significant contribution to the trajectory errors of the antenna phase center and the clock come from the position errors of the ISS center of mass.

In summary, the errors on the time transfer and on the relativistic correction have been expressed as functions of the trajectory knowledge through
equations (\ref{FinalEquationTimeTransfer}) and (\ref{RedshiftFinal}). Moreover, only the error in the knowledge of the ISS center of mass position
has an importance in the relativistic correction.

\section{Orbit error examples}
\label{orbito}

In this section we describe two simple models for ISS orbit errors as examples to investigate numerically the effect they have on ACES performance.


We consider two methods used to evaluate the orbit of a satellite. On the one hand, the dynamical method takes into account the equations of motion
to restitute the trajectory. It fits the measured data towards an orbit satisfying the principles of celestial mechanics and estimates a number of
parameters which give the best fit trajectory. On the other hand, the kinematic method uses directly the measurements of the satellite position (eg.
GPS data) with no a priori assumption of the form of the orbit. In both cases the resulting orbit errors are in general not easily described by a
simple model, but for our purposes we use two examples that should approximately reflect the orders of magnitude of the main orbit error
contributions.

For dynamical orbit determination the differences between true and computed trajectories of the ISS center of mass are expected to have specific
structures. For example an eccentricity error gives no long term effects, but periodic errors can be important and the radial, along track and
velocity errors are correlated. For weak eccentricity orbits, the difference between two real orbits is given by the Hill model (or the
Clohessy-Wiltshire model) which is an expansion of uncertainties with respect to a reference circular orbit \cite{Colombo,Colombo2}. If we suppose
there exists a set of parameters which perfectly describes the true orbit of the ISS, then this model is a first approximation to the errors of the
center of mass between the computed and the true orbits.

For kinematic orbit determination one expects little a priori correlation between the different orbit error components. We therefore use a simple
independent random noise on each component as a first approximation.


In the Hill model the position uncertainties along radial, tangential and normal axis (in the local ISS frame) are given as follows:

\begin{equation}
\begin{split}
  \text{radial axis :  } & \delta R(t) = \frac{1}{2} X \cdot  \cos (\Omega t + \varphi_R) + c_R \\
  \text{tangential axis :  } & \delta T(t) = - X  \cdot \sin (\Omega t + \varphi_R)
    - \frac{3}{2} \Omega \cdot c_R \cdot t + d_R \\
  \text{normal axis :  } &     \delta N(t) = Y \cdot \cos (\Omega t + \varphi_N) \\
  \label{RTNequations}
\end{split}
\end{equation}

where $X$, $Y$, $c_R$ and $d_R$ are amplitude coefficients, and where $\Omega$ is the orbital angular frequency. 
For our purpose, bias ($d_R$) plays no role and the linear term ($c_R$) depends on arc length. Basically the longer the observation duration is, the
smaller this coefficient becomes. The main feature of this error model is to take into account the error correlations in the orbital plane. For
instance, a positive radial bias leads to a negative error on the tangential velocity: the satellite is delayed with respect to the reference orbit.


\section{Numerical results}
\label{results}

We now use the previously described error models to calculate the corresponding constraints imposed by the ACES stability (equations
(\ref{BruitIntrinseque}) and (\ref{AttentesAuxTempsLongs})) and accuracy requirements, via equations (\ref{FinalEquationTimeTransfer}) and
(\ref{RedshiftFinal}). We first consider the Hill model, followed by results for a simple random noise.

We consider orbit error models in the local ($\overrightarrow{R}$,$\overrightarrow{T}$,$\overrightarrow{N}$) frame, so we need to transform them to
GCRS and determine the uncertainties in this frame on position and on velocity parameters. We consider an ephemeris of the ISS corresponding to the
$20^{th}$ of May, 2005 and a ground station based in Toulouse, France $(43^{o} 36' N, 1^{o} 26' E)$. Actually, this station has been chosen as the
master ground station of the ACES mission. All station parameters and their uncertainties have to be expressed in the same frame (ie. GCRS).

%
%

We first consider the error equation (\ref{FinalEquationTimeTransfer}) on the time transfer. We choose the signs of the independent parameters
($\overrightarrow{X_a X_a'}$, $T_{23}$ and $\delta T_{23}$) so as to maximize the resulting temporal Allan deviation. The calculated deviation has to
be compared with the MWL specifications.

Assuming we have no error on $T_{23}$ (ie. $\delta T_{23} =0 $ s), for all values of factor $X$ (or $Y$) of equation (\ref{RTNequations}), it is
possible to determine the maximum value of the time interval $T_{23}$ for which the temporal Allan deviation remains under the specifications. With
numerous values of $X$, we calculate a bound which marks out two different areas~: the allowed uncertainties area in which each couple ($X$,
$T_{23}$) gives a deviation staying under the specifications, and the prohibited area.
\begin{figure}[lh!]
\includegraphics[height=5cm]{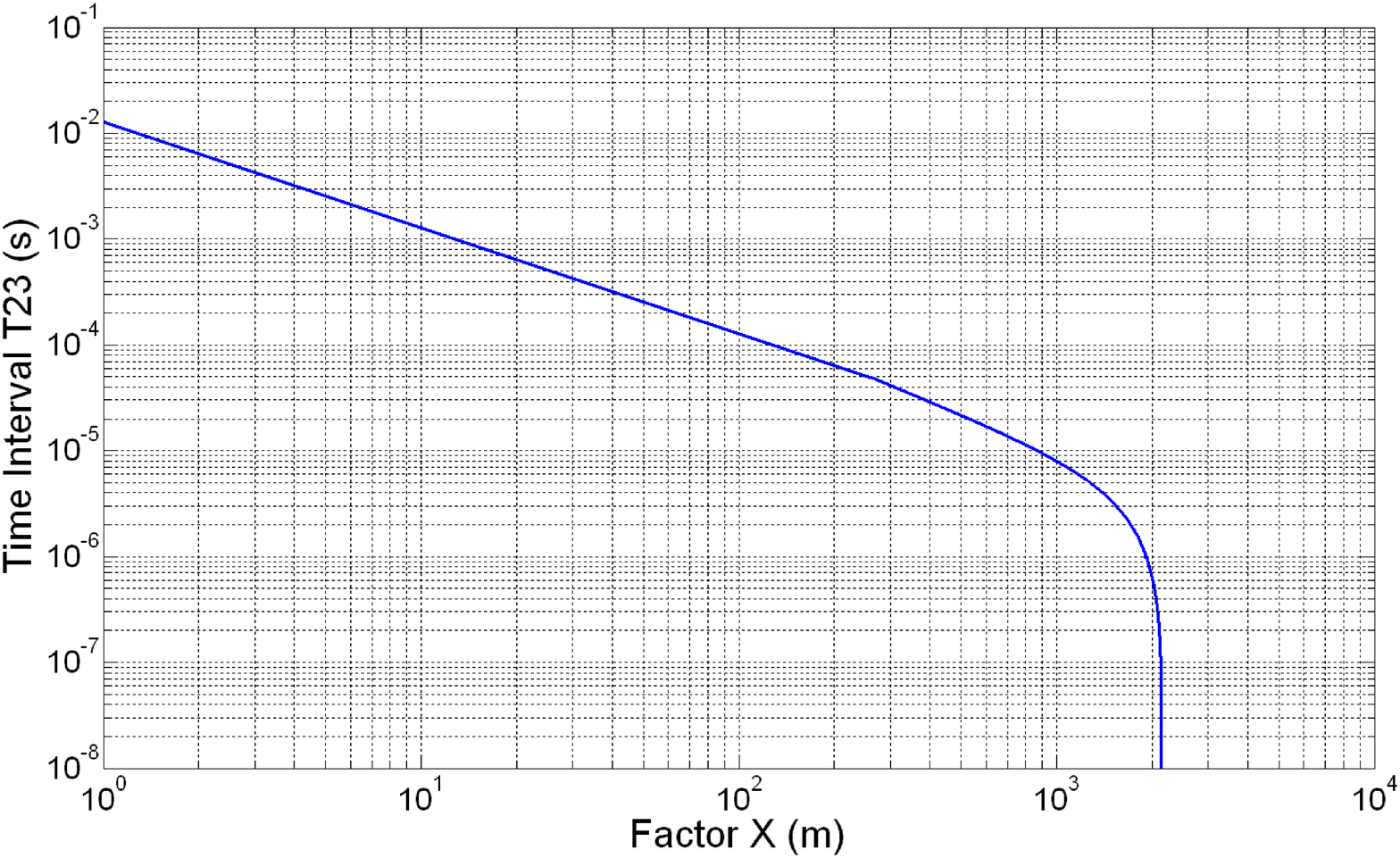}
    \caption{Maximum allowed value of $T_{23}$ as
    function of the scale factor $X$ to comply with the specifications, assuming $\delta T_{23} = 0$}
    \label{fig:T23fctfactorX}
\end{figure}

\begin{figure}[h!]
\includegraphics[height=3.6cm]{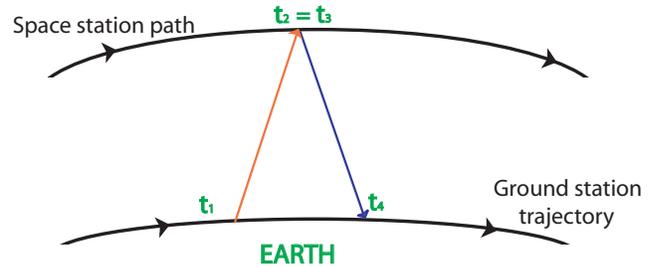}
    \caption{The "$\Lambda$ configuration" is the way to combine
     upwards and downwards signals which allows the maximum
     uncertainties on the space station orbit determination}
    \label{fig:ConfigurationLambda}
\end{figure}

Figure \ref{fig:T23fctfactorX} shows that, the smaller the time interval $T_{23}$, the larger the allowed uncertainty on the space station position.
This result provides a way to combine upwards and downwards signals to allow the maximum uncertainty on space station position in order to comply
with the specifications. The most favorable situation is when the reception at the antenna phase center of the space station corresponds to the
emission at the same place ie. $t_2 = t_3$ (see figure \ref{fig:ConfigurationLambda}). This way of combining signals is named the "$\Lambda$
configuration". To work with parameters in the asymptotic area (see figure \ref{fig:T23fctfactorX}) requires $T_{23}$ to be below $10^{-6}$ s.

\begin{figure}[h!]
\includegraphics[height=5cm]{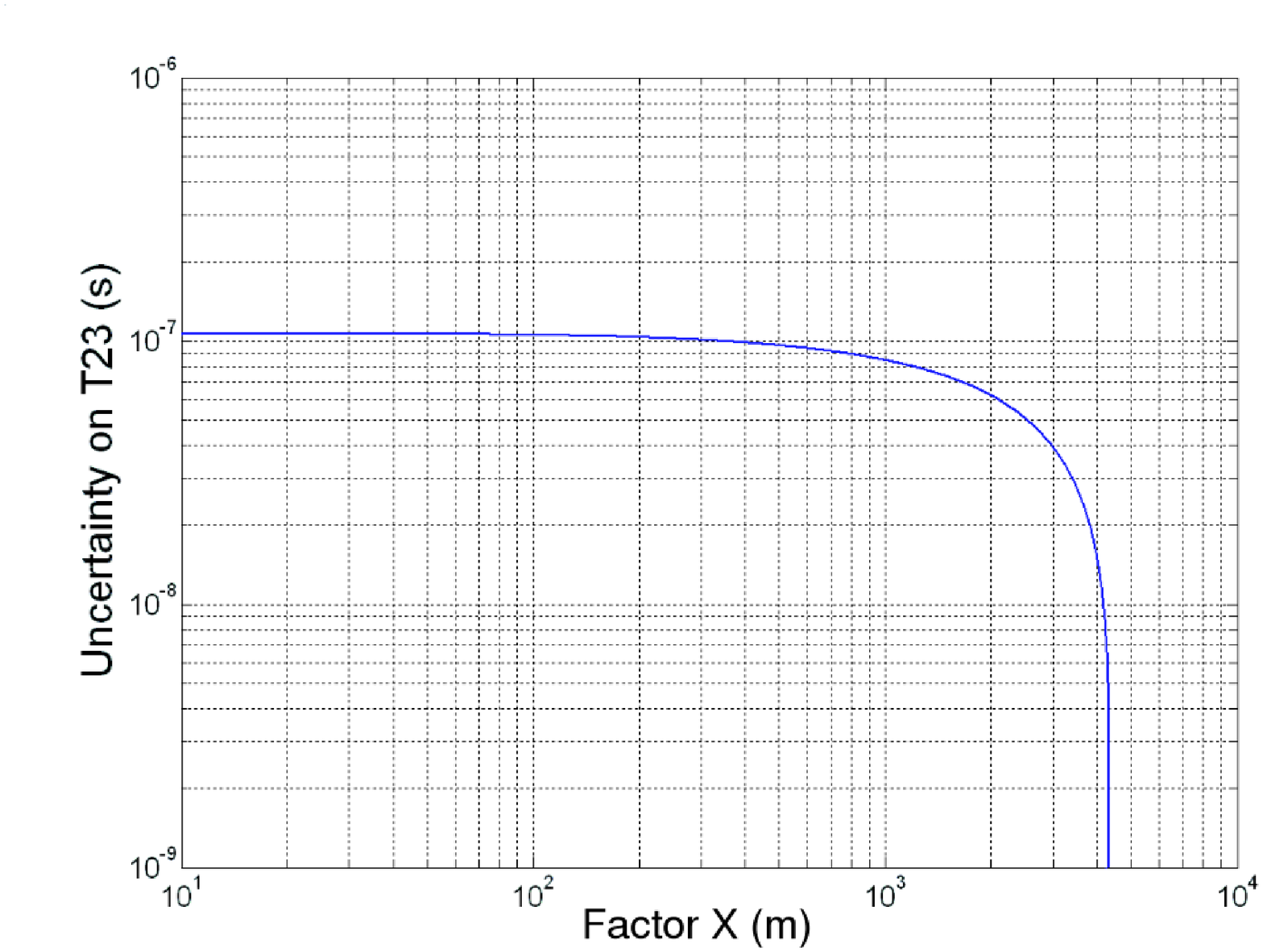}
    \caption{Maximum allowed value of $\delta T_{23}$ as
    function of the scale factor X to comply with the specifications, assuming $T_{23} = 0$}
    \label{fig:ArticleDeltaT23FctFactorX}
\end{figure}

Then if we plot the maximum value of $\delta T_{23}$ for all values of the factor $X$, there will appear two asymptotic values we cannot cross if we
want to stay within the specifications (see figure \ref{fig:ArticleDeltaT23FctFactorX}).
 Basically a compromise between the knowledge of
the space station trajectory and the precision of the internal
delays calibration must be achieved owing to the maximization of
the Allan deviation.  We will evaluate the maximum allowed errors
on these two parameters if no other errors are present.

We search for the asymptotic value of factors $X$ and $Y$ which comply with the specifications for all phases ($\varphi_R$, $\varphi_N$) when we have
no error on $T_{23}$. The asymptotic value for orbit determination is obtained for $X$, $Y$ = 4200 m which corresponds to a 4.2 km error on the
tangential and normal axes, and to a 2.1 km error on the radial axis.

\begin{figure}[h!]
\includegraphics[height=5cm]{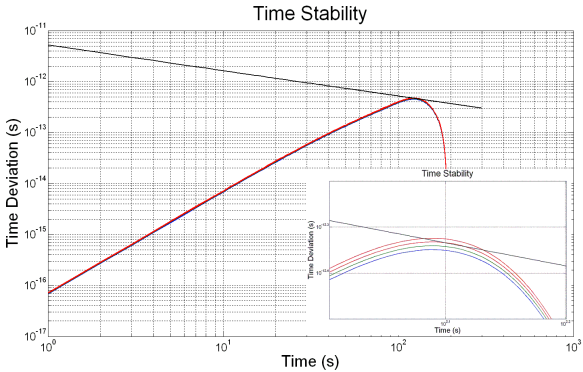}
    \caption{Temporal Allan deviations for $X$ = 0, $T_{23}= 0$ s and $\delta T_{23} = [102, 104, 106, 108]$ ns}
    \label{fig:AllNoisePhasesForDeltaT23&Factor=0}
\end{figure}

The asymptotic value of the time calibration does not depend on orbit determination uncertainties. So it is independent of the phases $\varphi_R$ and
$\varphi_N$. Then we can draw the temporal Allan deviations for different values of $\delta T_{23}$ (see figure
\ref{fig:AllNoisePhasesForDeltaT23&Factor=0}). We find that $\delta T_{23}$ must stay below à $1.06 \cdot 10^{-7}$~s.


The requirements for several passes have also been investigated. In this case, the calculated deviations have to be compared to the specifications
given by (\ref{AttentesAuxTempsLongs}). The results showed that the requirements on orbit determination and time calibration are less stringent for
several passes than for a single pass. Therefore if specifications are respected for a single pass, specifications for longer integration times are
also respected.

Now we evaluate requirements on orbit determination considering the relativistic frequency shift. We search for the maximum value of $X$ to comply
with the specifications (\ref{BruitIntrinseque}) and (\ref{AttentesAuxTempsLongs}). Equation (\ref{RedshiftFinal}) is evaluated with the error model
(\ref{RTNequations}), and its Allan deviation is calculated for different values of $X$. For integration times greater than one thousand seconds,
these Allan deviations are independent of the phases $\varphi_R$ and $\varphi_N$.
\begin{figure}[h!!]
\includegraphics[height=5cm]{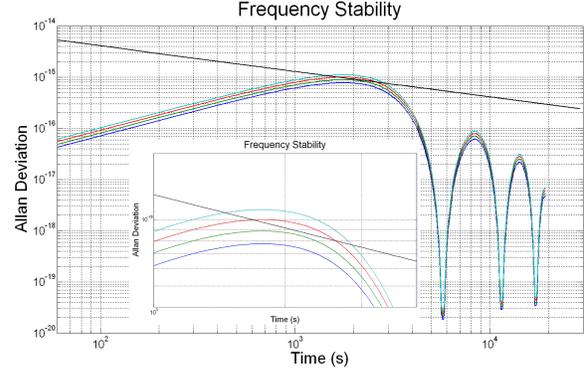}
\caption{Modified Allan deviations of the redshift error for $X$=14, 16, 18, 20 m} \label{fig:NewRedschiftRequirementsWithCorrelatedUncertainties}
\end{figure}

Figure \ref{fig:NewRedschiftRequirementsWithCorrelatedUncertainties} shows that, if the factor $X$ in (\ref{RTNequations}) is equal to 16 m ie. if we
have an eight meter error on the radial axis and sixteen meter error on the tangential axis, then we comply with the specifications.

Because of the projection of the position error along the ISS center of mass velocity (see equation (\ref{RedshiftFinal})), the requirement on the
factor $Y$ is orders of magnitude less stringent. The bound is given by the asymptotic value coming from the time transfer (see equation
(\ref{FinalEquationTimeTransfer}) and figure \ref{fig:ArticleDeltaT23FctFactorX}).

We now turn to the second example of orbit determination errors: independent random errors on $\overrightarrow{R}$, $\overrightarrow{T}$ and
$\overrightarrow{N}$. We consider white noise of amplitude (standard deviation) $\delta R$, $\delta T$, $\delta N$ at one second sampling intervals
on each of the three components.

Similarly to the Hill model, the by far more stringent constraints come from the effect of these errors on the relativistic frequency shift via
equation (\ref{RedshiftFinal}). In particular, a white noise $\delta T$ in equation (\ref{RedshiftFinal}) will translate into a temporal Allan
variance given by
\begin{equation}
\sigma_x (\tau) \simeq \frac{V_0 \delta T}{c^2} \cdot \tau^{-\frac{1}{2}}, \label{BruitenT}
\end{equation}
which satisfies the requirements (\ref{BruitIntrinseque}) and (\ref{AttentesAuxTempsLongs}) for all integration times when $\delta T \leq 60$~m.

Because of the projection of the error on $\overrightarrow{V_0}$ in (\ref{RedshiftFinal}) the relativistic frequency shift imposes virtually no
limits on the normal and radial components of the errors. Those are limited by the effect on the time transfer given by equation
(\ref{FinalEquationTimeTransfer}). Assuming $T_{23}\leq 10^{-6}$~s and $\delta T_{23} \leq 10^{-7}$~s (see above) the first term of that equation is
by far dominant. Maximizing the dot product in that term we obtain an upper limit of 1.4~km on $\delta R$ and $\delta N$ in order to stay within
specifications for all integration times.

Finally, we consider the accuracy requirement of ACES i.e. $10^{-16}$ in relative frequency when averaged over ten days. From the integral of
equation (\ref{RedshiftFinal}) this implies that the tangential component of the position error ($\overrightarrow{X_c X_c'}$ in
(\ref{RedshiftFinal})) cumulated over ten days needs to remain below one kilometer (including for example the linear term along the tangential axis
in (\ref{RTNequations})). This is unlikely to raise any difficulty, if the much more stringent requirements from periodic or random errors (see
above) are met. We note that the accuracy requirement is related to a requirement on the knowledge and conservation of the total energy of the orbit.
An error in the total energy will show up as a velocity bias, and thus as a linear term along the tangential axis. The non-conservation of the total
energy (non-gravitational accelerations) was found to be negligible in section \ref{Effects}. Nonetheless possible long-term effects, if they exist,
will lead to a slow variation of the tangential error. The constraint on both, the error in the total energy and the effects due to energy
non-conservation is provided by the accuracy requirement, i.e. that the tangential error cumulated over ten days needs to remain below one kilometer.

%
%


\section{Conclusion}
\label{conclusion}

We have derived detailed and general (within the specified assumptions) expressions for the calculation of the effect of orbit determination and
instrumental calibration errors on the stability and accuracy of ground to space clock comparisons. These expressions were then used on the specific
example of the ACES mission to derive maximum allowed errors given the stability requirements of that mission. For that purpose we have used two
simplified orbit error models (Hill model, and random noise), setting limits on the appropriate parameters of those models. Although neither of those
models is likely to correctly reflect all of the ISS orbit errors, they are nonetheless expected to provide correct orders of magnitude for the
maximum allowed orbit determination uncertainties. They are summarized in table \ref{TableResults}, where in each column we have provided the more
stringent result from the two models.

Furthermore, we have identified an optimal way to combine upwards and downwards signals (the $\Lambda$ configuration), which allows for the maximum
orbit determination errors. This then provides a constraint on the maximum allowed value of $T_{23} \leq 10^{-6}$~s and its maximum allowed
uncertainty of $\delta T_{23} \leq 10^{-7}$~s, which constrains the required knowledge of the sum (transmission + reception) of instrumental delays
between the antenna phase centre and the onboard clock (also provided in table \ref{TableResults}).

\begin{table}[h!!]
\caption{Requirements on orbit determination and internal delay calibration} \centering
\begin{tabular}{c|c|c|c}
  $\delta R$ /m & $\delta T$ /m & $\delta N$ /m & $\delta T_{23}$ /s \\
  \hline
  8 & 16 & 1400 & $10^{-7}$ \\
\end{tabular}
 \label{TableResults}
\end{table}

Finally, the accuracy requirement of ACES was used to constrain long term ($\approx 10$~days) linear drifts and other slow variations of the orbit
determination errors. We find that the total tangential error cumulated over 10 days needs to stay below 1~km to comply with the accuracy
specification ($10^{-16}$ in relative frequency after ten days averaging), which should pose no particular problems given the more stringent short
term requirements (see table \ref{TableResults}).

In conclusion the requirements on orbit determination are significantly less stringent than the initial 'naive' estimate (one meter error for
$10^{-16}$ in relative frequency), which is mainly due to partial cancelation between the gravitational redshift and the second order Doppler effect
in the relativistic frequency correction of the onboard clock.

%
%

\begin{acknowledgements}
The research project is supported by the French space agency CNES
and ESA through Lo\"\i c Duchayne's research scholarship $N^{o}$
05/0812.
\end{acknowledgements}




\end{document}